\documentclass[a4paper,12pt]{article}

\usepackage{graphicx}
\usepackage{dcolumn}
\usepackage{bm}
\usepackage{amsmath}
\usepackage{setspace}
\usepackage[super,comma,sort&compress]{natbib}

\newcommand{\UNIT}[1]{\ensuremath{\:{\rm #1}}}
\newcommand{\CHEM}[1]{\ensuremath{\mathrm{#1}}}

\newcommand{\glas}[2]{\ensuremath{\mathrm{#1S\!\!-\!\!#2}}}

\newcommand{\ns}{\UNIT{ns}}
\newcommand{\ps}{\UNIT{ps}}
\newcommand{\fs}{\UNIT{fs}}
\newcommand{\eV}{\ensuremath{\:e\mathrm{V}}}

\begin{document}

\singlespacing

{\centering
{\Large
The cationic energy landscape in alkali silicate glasses: properties and relevance
}

\bigskip
\bigskip

Heiko Lammert

hlammert@ucsd.edu

\medskip

Center for Theoretical Biological Physics,
University of California San Diego,
9500 Gilman Dr., La Jolla, CA 92093-0374

\bigskip

Radha D. Banhatti

banhatt@uni-muenster.de

\medskip

Andreas Heuer

andheuer@uni-muenster.de

\medskip

Institute of Physical Chemistry and Sonderforschungsbereich 458,
University of M\protect\"unster, Corrensstr. 30,
D-48149 M\protect\"unster, Germany

\bigskip

}

\section*{Abstract}

Individual cationic site--energies are explicitly determined from
molecular dynamics simulations of alkali silicate glasses,
and the properties and relevance of this local energetics to ion transport
are studied.
The absence of relaxations on the timescale of ion transport
proves the validity of a static description of the energy landscape,
as it is generally used in hopping models.
The Coulomb interaction among the cations turns out essential to obtain an
average energy landscape in agreement with typical simplified hopping models.
Strong correlations exist both between neighboring sites and between
different energetic contributions at one site,
and they shape essential characteristics of the energy landscape.
A model energy landscape with a single vacancy is used to demonstrate
why average site--energies, including the full Coulomb interaction,
are still insufficient to describe the site population of ions, or their
dynamics.
This model explains how the relationship between energetics and ion dynamics
is weakened, and thus establishes conclusively that a hopping picture with
static energies fails to capture all the relevant information.
It is therefore suggested that alternative simplified models of ion
conduction are needed.

\newpage



\section{Introduction}

The transport of mobile modifier ions in silicate glasses below the
glass transition takes place in a basically fixed environment
provided by the glass structure. Besides the disordered structure of
the network, the Coulomb interactions among the mobile ions add
significantly to the complexity of the underlying energy landscape,
because their contribution is both long--ranged and non--static.

A hopping picture of the dynamics has been widely
accepted,\cite{baranovskii:1999} based on evidence for the
existence of well defined ionic sites, from
experiments\cite{greaves:1981,greaves:1985,rouse:1978,kamijo:1996,swenson:2001a}
and also from
simulations\cite{balasubramanian:1993,habasaki:1995,park:1999}.
Basically hopping models describe the glass system by a
representation of the effective energy landscape experienced by
the ions. The conformational disorder of the network is typically
represented by a distribution of site--energies.
They are treated as static, relying on the separation of timescales
between ion dynamics and structural relaxation.\cite{angell:1983}
Energy values for individual sites and
barriers are normally treated as independent of each other in
realizations by lattice models; see, e.g.,
\cite{baranovskii:1999}. This common assumption is however not
trivial, and it has been dropped in at least one
case.\cite{peibst:2005} Numerical investigations of such
simplified models have established that the Coulomb interaction
between the mobile ions is required for a completely correct
representation of the dynamics,\cite{maass:1999} and a
quantitative model of the ionic conductivity has been built with
the interactions among the ions as the main
element.\cite{funke:2004} Most other approaches however disregard the
Coulomb interaction among the mobile
ions.\cite{bunde:1991:2,habasaki:1995,hunt:1997,maass:1999,kirchheim:2001,swenson:2001a,ingram:2006,imre:2006}
This immediately raises the question whether the Coulomb
interaction can be simply taken into account by a local
modification of the energy landscape.

Recently a statistical method allowed us the identification of all
ionic sites in realistic molecular dynamics $(\mbox{MD})$
simulations.\cite{lammert:2003} Most of the conclusions have been
confirmed by other groups. \cite{vogel:2004, habasaki:2004}. By
direct observation of the ion hopping dynamics, key predictions
about the ion dynamics
could be tested.\cite{lammert:2005}. Furthermore, different spatial aspects
of the conduction paths have been reported
\cite{jund:2001,meyer:2002,horbach:2002,adams:2002,habasaki:2007}.

Most of the hopping models mentioned above use the concept of
site--energies. However, typically the properties of these site--energies
are postulated in an ad-hoc manner. In the present work we express
the properties of the energy landscape via explicitly determined
site-energies. These are obtained based on our $(\mbox{MD})$ simulations
of alkali silicate systems, from time-averages of the particle energy of
the ion residing in a site.
Two key questions are addressed:
What are the properties of the site--energies for a real
ion conductor? Do these site--energies indeed determine
thermodynamic or dynamic properties of the corresponding sites?
In particular, since the Coulomb interaction between cations can be expected
to be strong, it is not evident a priori that it can be taken into account as
merely a contribution to single-particle or, equivalently, single--site
energies. This aspect is discussed in detail by examining the correlation
of site--energies to thermodynamical and dynamical observables.

\section{Method}

In our MD simulations the potential energy is given by pair
interactions as $U_{\mathrm{sys}} = 1/2 \sum_{i,j\neq{}i} U_{ij}$.
The interaction potential $U_{ij}(r_{ij}) = q_i q_j /
(4\pi\varepsilon_0 r_{ij}) -A_{ij} r_{ij}^{-6} + B_{ij}
\exp(-C_{ij} r_{ij})$ consists of the Coulomb interaction for
point charges $U^{\mathrm{Q}}_{ij}$ and of a short range part
$U^{\mathrm{BM}}_{ij}$ of Born-Mayer type. The $q_{i,j}$ reflect
appropriately chosen partial charges \cite{habasaki:1995}.

For a finite system, single-particle energies $U_i$ can be computed
just like $U_{\mathrm{sys}}$ as the sum of all pair contributions
$U_{ij}$ involving the particle $i$, $U_i = \sum_{j \neq i} U_{ij}$.
While $\sum_i U_i = 2 U_{\mathrm{sys}}$, any change $\Delta
U_{\mathrm{sys}}$ to the energy of the system resulting from a
change of particle $i$ is directly given by $\Delta U_i$.
Considering a cation as particle $i$, the contribution to $U_i$ only
from interactions with e.g. other cations can furthermore be
isolated simply by restricting the sum to the appropriate $j$.

The situation is complicated by the use of periodic boundary
conditions: To avoid finite-size effects, the system is treated as
an infinite repetition of periodic images of the simulation
box.\cite{frenkel:2002} Due to the long range nature of the Coulomb
interaction, it cannot be computed by simple summation up to a
finite cutoff in this case. Instead all periodic images of the
simulation box must be taken into account. We use the Ewald
method\cite{frenkel:2002} for this purpose, where the Coulomb energy
is broken into three formal contributions: Only the real part
$U^{\mathrm{RE}}_{ij}$ is a pair term like $U^{\mathrm{BM}}_{ij}$;
the Fourier part $U^{\mathrm{FO}}_{ij}$ relates to the infinite set
of periodic images of particle $j$, and the self-correction
$U^{\mathrm{SE}}_i$ is globally assigned to particle $i$ only.

Yet a deeper analysis shows that the Coulomb part $U^{\mathrm{Q}}_i$
of a single-particle energy can still be assembled from these terms,
and that it can still be divided into contributions from groups of
like particles. One can notably
still separate the interaction of a cation with other cations
$U^{\mathrm{cat}}_i$ from the interaction with network atoms
$U^{\mathrm{net}}_i$, because $U^{\mathrm{SE}}_i$ can be traced
exclusively to the cation part. A constant
correction\cite{figueirido:1995} for the nonzero charge of the
partial systems under consideration is neglected, because it does
not affect the results of this study, where energy differences are
relevant.

As the locations and structural properties of the sites are stable
on the time scale of alkali ion diffusion, the potential energy of
cation $i$ is determined by the site $s(i)$ it occupies. A
site--energy $E(s)$ can therefore be identified by taking the time
average over the particle energy of the ions $i(s,t)$ residing in
$s$ at any given time $t$, i.e. $E(s) = \left< U_{i(s,t)}(t)
\right>_{t}$. Like the particle energies, the total value of the
site--energy $E^{\mathrm{tot}}(s)$ is separated into the
contributions $E^{\mathrm{cat}}(s)$ and $E^{\mathrm{net}}(s)$. Of
course, with this construction $E(s)$ cannot be interpreted as a
bare site--energy but reflects the site--energy {\it under the
condition} that it is populated by an ion.

\section{Systems}

We analyzed the site--energies for the alkali disilicate systems
$(\CHEM{Li}_2\CHEM{O})\cdot2(\CHEM{SiO}_2)$ and
$(\CHEM{K}_2\CHEM{O})\cdot2(\CHEM{SiO}_2)$, named \glas{L}{2} and
\glas{K}{2} respectively. These systems have already been
described in our previous work\cite{lammert:2005}. They contain
270 cations among 1215 atoms, at experimental
densities\cite{bansal:1986}. Interactions are given by a potential
by Habasaki et al.\cite{habasaki:1992}. The systems were
propagated  with a timestep of 2 \fs{} in the canonical ensemble,
using a Nos\'{e}-Hoover thermostat\cite{hoover:1985}. Positions
and energy contributions were stored every 0.1 \ps, with energies
averaged over five values sampled 20 \fs{} apart. Energy data was
produced for 10 \ns{} of simulations at 850 \UNIT{K}.

Our analysis identifies for these datasets 291 and 276 sites
respectively in \glas{L}{2} and \glas{K}{2}. Ions are in all cases
residing in a site during more than $98\%$ of the times, according
to our algorithm that drops excursions.\cite{lammert:2003}

\section{Results}

\subsection{Properties of average site--energies}

It turns out that the fundamental characteristics of the energy
landscape are qualitatively identical in both systems. Note that
these energies result from long time averages. Possible temporal
fluctuations are discussed further below. Broad ranges of
site--energies reflect the variation in the disordered structure.
For $E^{\mathrm{tot}}$ the distributions can be fitted by
Gaussians, with the parameters given in Tab. \ref{tab_totfit}.
The properties of the
distributions are compatible with earlier results, determined both
for sites\cite{habasaki:2004}, and without reference to ion
sites\cite{balasubramanian:1995,habasaki:1996,park:1999,sunyer:2002}.

\begin{table}
\begin{center}
\begin{tabular}{lcc}
System     &    $\mu [\eV]$ & $\sigma_{dist} [\eV]$ \\
\glas{L}{2}     & -8.8  & 0.33     \\
\glas{K}{2}     & -5.6  & 0.24     \\
\end{tabular}
\caption{\label{tab_totfit}Gaussian fits for the distribution of site energies
$f( E^{\mathrm{tot}})$ }
\end{center}
\end{table}

\begin{figure}
\begin{center}
\includegraphics[clip,width=8.6cm]{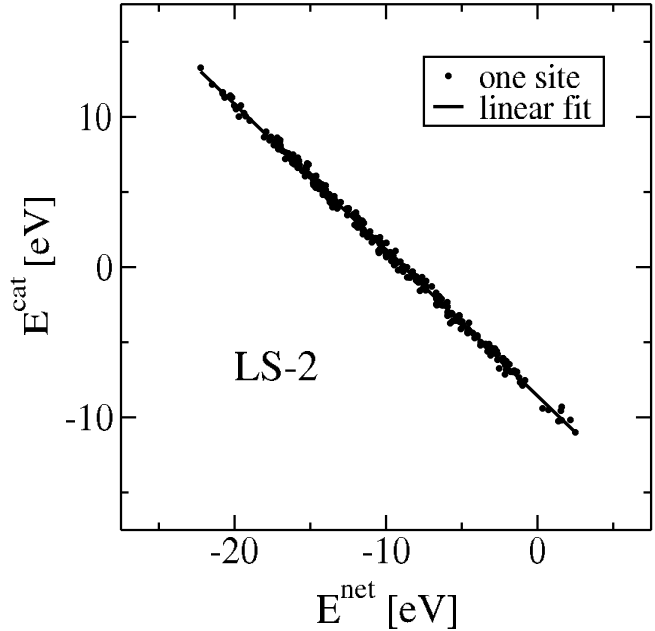}
\caption{\label{fig_catnet} Contributions $E^{\mathrm{cat}}(s)$ and
$E^{\mathrm{net}}(s)$ for each site $s$ in system \glas{L}{2} }
\end{center}
\end{figure}

The influence of interactions with both cations and network on the
shape of the site--energy distributions is demonstrated in Fig.
\ref{fig_catnet} for the system \glas{L}{2}. For each site the
cation contribution $E^{\mathrm{cat}}(s)$ is plotted against the
corresponding network energy $E^{\mathrm{net}}(s)$. Both sets of
energies have standard deviations of 5 to 6 \eV{} and cover ranges
of more than 20 \eV. In the system \glas{K}{2} standard deviations
of 3 \eV{} and ranges of 16 \eV{} are found.
In all cases the significantly narrower distributions for
$E^{\mathrm{tot}}(s)$ are reached because a strong and clear
anticorrelation between $E^{\mathrm{cat}}(s)$ and
$E^{\mathrm{net}}(s)$, evident in Fig. \ref{fig_catnet}, reduces
the spread of values for the resulting total site--energies.

Naturally, the strong variation of $E^{\mathrm{net}}$ can be
interpreted as a consequence of the disordered network structure.
The presence of non-bridging oxygens (NBO) gives rise to a negative
partial charge. Thus, due to the strong Coulomb interaction, a
minor variation in the distance between the site and the nearest
NBO may result in a significant variation of the network energy.
The typical distance between a \CHEM{Li}-ion and a NBO is 2\AA.
For example, variation of this distance by 0.1\AA{} gives rise to
an energy variation of $\pm 0.4 \eV{}$, for the potential
used in these simulations.

\begin{figure}
\begin{center}
\includegraphics[clip,width=8.6cm]{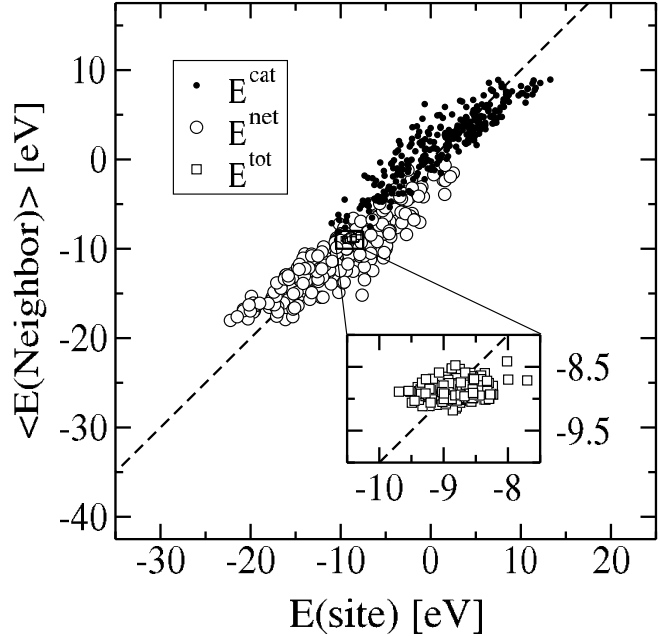}
\caption{\label{fig_eneigh} Correlation among the energies of
neighboring sites in system \glas{L}{2} }
\end{center}
\end{figure}

To test the assumption of many lattice models that the
site--energies are uncorrelated, the energies of all sites were
compared to those of their next neighbors, defined by the first
nearest-neighbor shell of the alkali pair distribution. The
results are shown in Fig. \ref{fig_eneigh} for the glass
\glas{L}{2}, whose behavior is again typical for all investigated
systems. For each site $s$ the average energy of the neighbors of
$s$ is plotted against the site--energy of $s$ itself. Data for
the contribution $E^{\mathrm{net}}$ of the matrix alone is given
by the open circles. They show a strong spatial correlation of the
network energies between adjacent sites. The same correlation is
found for the cation interactions $E^{\mathrm{cat}}$, as it may be
expected when $E^{\mathrm{cat}}$ and $E^{\mathrm{net}}$ are
anticorrelated themselves. The total energies are shown as squares
in the inset. For them spatial correlations are much weaker (regression coefficient
0.26 $\pm$ 0.1),
because the remaining imperfections in the cancelation of
$E^{\mathrm{cat}}$ and $E^{\mathrm{net}}$ introduce sufficient
fluctuations into $E^{\mathrm{tot}}$ to create a nearly uncorrelated
disordered landscape.

\subsection{Temporal variations of site--energies}

Temporal variations of site--energies can occur as a consequence of
different physical effects: (1) Gradual structural relaxation of
the silicate glass, (2) Fluctuations due to motion of adjacent ions. Both
aspects are analyzed in this part.

The long-time variation of $E(s)$, resulting from the structural relaxation, was estimated by comparing the
mean energy values for each site from the first and from the
second half of the available $10 \ns$ of data. Judging from the
width of the Gaussian distribution of these energy differences,
the residual long-time fluctuations of a site--energy
$E^{\mathrm{tot}}_{\CHEM{Li}}$ are $0.03 \eV$ and $0.02 \eV$
for the Lithium and Potassium glasses.
These fluctuations, related to {\it systematic} drifts are small as
compared to the overall width of the Gaussian distributions; see
Tab. I.

As mentioned above the energy value for a single site results from
an average over the whole simulation run.
Random fluctuations in this site-energy can be observed by
defining the energy value instead from very short averages.
Two time scales have to be compared. First,
the time scale $\tau_{fluct}$ characterizes the time where typical
energy fluctuations of a site are explored. Second, $\tau_{res}$
describes typical residence times of an ion in a site. The
distribution of $\tau_{res}$ has a median of a few hundred
picoseconds (see below). Only if $\tau_{fluct} \ll \tau_{res}$ it
is justified to characterize the properties of a site, as
experienced by an ion, via the average $E(s)$.  In the opposite limit one would have to introduce a time--dependent energy $E(s)$.

\begin{figure}
\begin{center}
\includegraphics[clip,width=8.6cm]{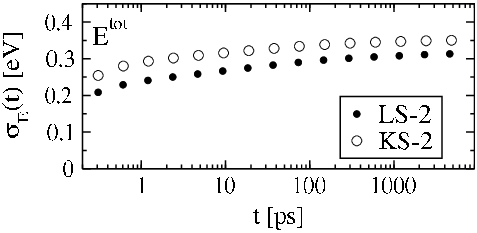}
\includegraphics[clip,width=8.6cm]{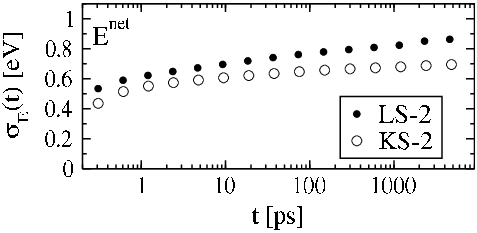}
\caption{\label{fig_sigmae} Growth of the standard deviation
$\sigma_E$ of energies for a site with sampling time. The behavior of
$E^{\mathrm{cat}}$ is similar to that for $E^{\mathrm{net}}$. }
\end{center}
\end{figure}

In order to check how stable and distinguishable the individual
site--energies are, we plot in Fig. \ref{fig_sigmae} the average
growth of the standard deviation $\sigma_E(t)$ for the set of
energy values that are incorporated into the mean site--energy.
The variable $t$ denotes the width of the interval, used for
calculating the average.
The development of the
total energy $E^{\mathrm{tot}}$ is shown in the upper part, the
lower part treats the network contribution $E^{\mathrm{net}}$. The
cation part $E^{\mathrm{cat}}$ is omitted, because its behavior is
very similar to that of $E^{\mathrm{net}}$, due to the correlation
demonstrated above.

First, it turns out that for \glas{L}{2} beyond 100 ps a nearly constant
value of $\sigma_E(t)$ is observed. Thus, indeed, the typical
fluctuation time $\tau_{fluct}$ is much shorter than $\tau_{res}$.
The range of values sampled for the energy of a typical site has a
considerable width.
It reaches a value of $\sigma(t=4.6\ns) = 0.31 \eV$
at the end of the plot for $E^{\mathrm{tot}}_{\CHEM{Li}}$ in \glas{L}{2},
and of $\sigma(t=4.6\ns) = 0.35 \eV$
for $E^{\mathrm{tot}}_{\CHEM{K}}$ in \glas{K}{2}.
The overall stability of the energies on a timescale of more
than 5 \ns{} is supported by the small errors for $E^{\mathrm{tot}}$
determined above.

The site--energies, introduced so far, provide an average view of
the energy landscape, as it is adopted by most hopping models.
In principle one might envisage a scenario where dynamical events
induce systematic changes in the local configuration. Specifically the energy
of a site might change as a consequence of an ionic jump. For
example, in the MIGRATION concept \cite{funke:2004} an ion,
jumping to a site, is thought to be gradually stabilized by the subsequent
adaption of the other ions to that jump. In this scenario one
might expect a gradual decrease of the site--energy via a decrease
of $E^{\mathrm{cat}}$. Of course, for reasons of time-reversal
symmetry it has to increase correspondingly before the jump to the
next site.

\begin{figure}
\begin{center}
\includegraphics[clip,width=8.6cm]{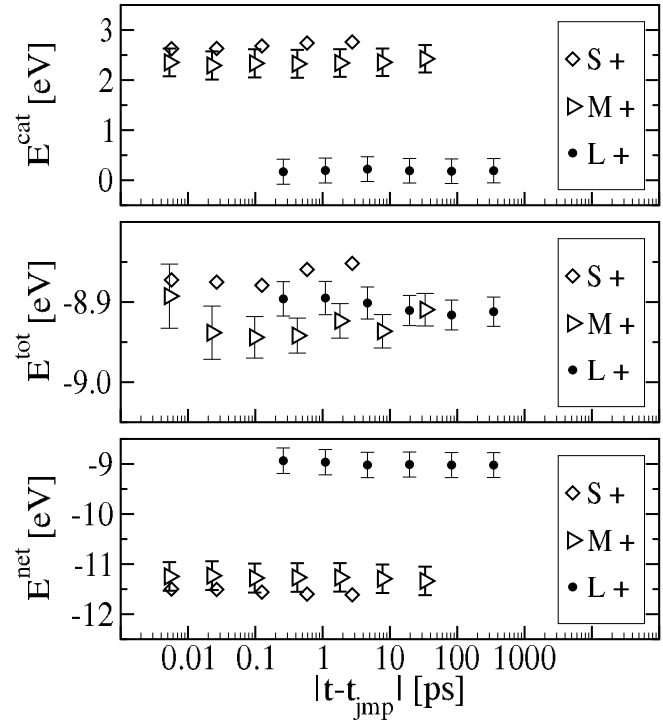}
\caption{\label{fig_etime}Development of the energy of an ion
directly after the jump into a site, at $t=t_{\mathrm{jmp}}$.
Three sets of residences with different duration are distinguished,
designated Short, Medium and Long.
Behavior at the end of a residence before the jump is equivalent.}
\end{center}
\end{figure}

To analyse a possible time-dependence we recorded the
time-dependence of the energy during residences by the ions.
Here we restrict ourselves to the system \glas{L}{2}.  A possible
time-dependence would naturally depend on the typical residence
time of an ion in a site. Therefore, we have grouped the different
residence times into bins, namely S: $\tau \in [10 \ps{},100\ps{}
]$, M: $\tau \in [100 \ps{},1 \ns{}]$, and L: $\tau \in [1 \ns{},
10\ns{}]$. The results are shown in Fig. \ref{fig_etime}. For
groups S and M we performed additional simulations with a sampling
interval of 2 \fs{} (during 1 ns). For group L the data from our
main simulations is shown, with a sampling interval of 100 \fs{}.
Plotted times extend up to 10 \% of the maximum duration, such
that values are available from all residences in the group.

The time dependencies of the energies from the start and
from the end of the residence were both investigated. The positive
direction of time was defined in both cases away from the nearest
jump, such that the plotted time increases always towards the
middle of the residence. Only the development from the start of
the residences is shown here. In all cases the data from the end
of the residences is identical within the errors, given by either
error bars or symbol size.

It is clearly seen that, if at all, $E^{\mathrm{cat}}$ and
$E^{\mathrm{net}}$ slightly change with time for the shortest
residences. After their addition to the total energy
$E^{\mathrm{tot}}$ only unsystematic fluctuations remain. In
conclusion no energetic relaxations are observed when an ion
reaches a site, or before it leaves. The average site--energies
are therefore the appropriate quantities for a deeper analysis of the
behavior of the systems.

\subsection{Site--energy vs. thermodynamics}

The residence times are linked to the site--energies in an
indirect way, via the height of the energy barrier between two
sites. While it is probably not crucial, this complication is
absent for the thermodynamic occupation probabilities of the
sites, which should depend directly on their energies via a Fermi
distribution.

\begin{figure}
\begin{center}
\includegraphics[clip,width=8.6cm]{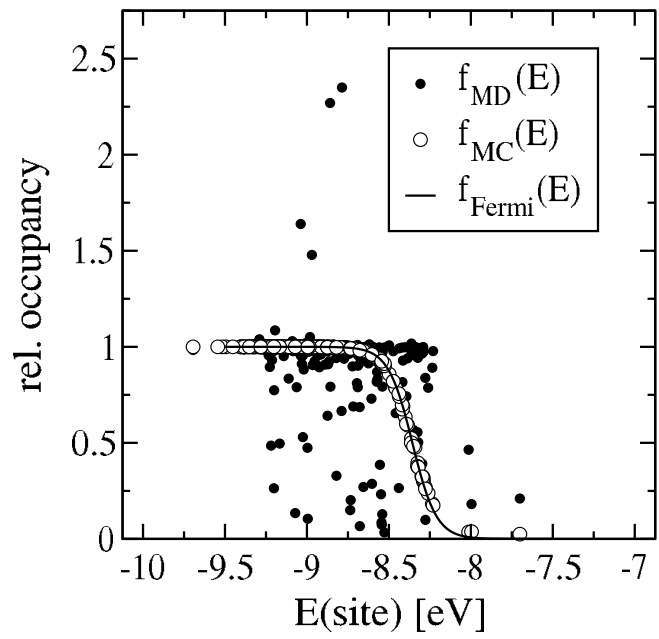}
\caption{\label{fig_fermi}
Mean relative occupation of the sites in \glas{L}{2}
plotted against $E^{\mathrm{tot}}$.
The solid line is a Fermi-Dirac function with $\mu = 8.4 \eV{}$,
the open symbols were generated from a Monte Carlo model that recreates
the MD system (see text).
}
\end{center}
\end{figure}

In Fig. \ref{fig_fermi} this prediction is tested for the system
\glas{L}{2}. A Fermi function with $\mu = -8.4 \eV{}$ is given by
the solid line. The value of $\mu$ was chosen such that the
predicted total occupation of all sites from the $\mbox{MD}$
simulation corresponds to the number of alkali ions. However the
occupation data for the individual sites from the $\mbox{MD}$
simulation, given by the solid circles, deviate fundamentally from
a Fermi function. No clear dependence of the relative occupation
on the site--energy can be observed. For comparison, a system with
the same number of sites and ions and with the site--energies
$E^{\mathrm{tot}}$ found in the $\mbox{MD}$ system was propagated
in a Monte Carlo simulation. The occupation data from this
simulation, given by the open circles, nicely agrees with the
theoretical Fermi function.

Some minor deviations in the MD data might occur from the fact that in
disagreement with the derivation of the Fermi relationship\cite{gusev:1991}
many sites may very rarely host two ions at
the same time. However, as seen from Fig. \ref{fig_fermi} only 4
sites have an average number of hosted ions which is significantly
larger than unity. Therefore this effect cannot be responsible for the
significant deviations.

\subsection{Site--energy vs. dynamics}

\begin{figure}
\begin{center}
\includegraphics[clip,width=8.6cm]{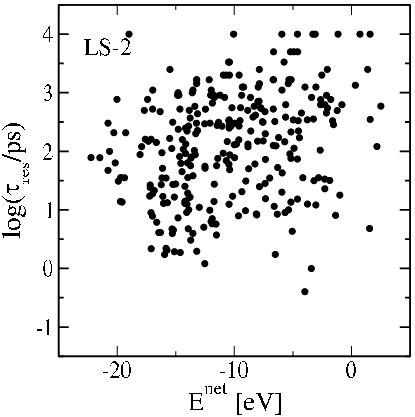}
\includegraphics[clip,width=8.6cm]{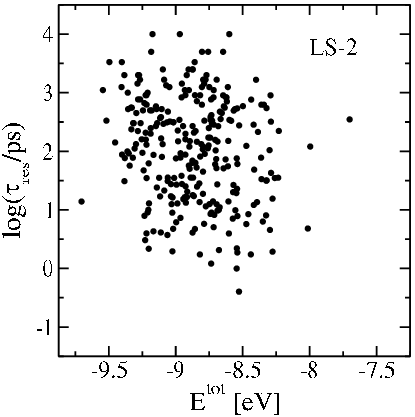}
\caption{\label{fig_timefit} Mean residence times
$\tau_{\mathrm{res}}$ plotted against the site--energies
$E^{\mathrm{net}}$ and $E^{\mathrm{tot}}$. }
\end{center}
\end{figure}

Another interesting observation in Fig.\ref{fig_etime} is the
correlation of $E^{\mathrm{cat}}$ and $E^{\mathrm{net}}$ with the
residence time. For long residences, represented by $L$,
$E^{\mathrm{cat}}$ is smaller whereas
$E^{\mathrm{net}}$ displays the opposite trend.
The data in Fig. \ref{fig_etime} also suggest that a minor correlation
along this line seems to remain for $E^{\mathrm{tot}}$. Sites with
short residence times have somewhat higher total energies
$E^{\mathrm{tot}}$. This has been systematically analyzed in Fig.
\ref{fig_timefit} for the system \glas{L}{2}. For each site the
mean residence time $\tau_{\mathrm{res}}$for ions in a site is
plotted against the network energy $E^{\mathrm{net}}$ and
against the site--energy $E^{\mathrm{tot}}$.

On average there is some increase of the residence time
with increasing network energy. As directly indicated by the major
scatter the correlation coefficient between $\log
\tau_{\mathrm{res}}$ and $E^{\mathrm{net}}$ is relatively small
(0.28). The dependence of $\log \tau_{\mathrm{res}}$ on $E^{\mathrm{tot}}$
is also present, but shows an even weaker correlation coefficient.(-0.18)
A possible reason for these weak correlations is discussed below.

\section{Model energy landscape}

The deviations of ion behavior from expectations based on the average energies,
like the Fermi distribution for relative occupations, can be
understood when considering the effect of ion-ion interactions.
To visualize this
effect we use a very simple model with $N$ sites and $N-1$ ions.
This choice reflects the observed small number of free sites. Let
$E^{\mathrm{net}}(s)$ denote the energy of an ion at site $s$ due
to the interaction with the network.  Furthermore,
$E^{\mathrm{vac},t}(s)$ denotes the additional energy of an ion at
site $s$ due to the interaction with the other ions under the
condition that site $t$ ($t \ne s$) is empty. One expects a small
but significant dependence of $E^{\mathrm{vac},t}(s)$ on $t$,
e.g., via the distance of site $t$ to site $s$. Furthermore, this
term also contains possible correction effects of the network
energy because the network structure may also depend on the actual
ionic configuration. The energy $E^{\mathrm{vac,total}}(t) =
\sum_{s \ne t} [E^{\mathrm{net}}(s) + E^{\mathrm{vac},t}(s)] $
thus denotes the total energy of the system under the condition
that site $t$ is vacant. Correspondingly, the equilibrium
probability $p^{\mathrm{vac},t}$ that site $t$ is empty is
proportional to $\exp(-\beta E^{\mathrm{vac,total}}(t))$.

\begin{figure}
\begin{center}
\includegraphics[clip,width=8.6cm]{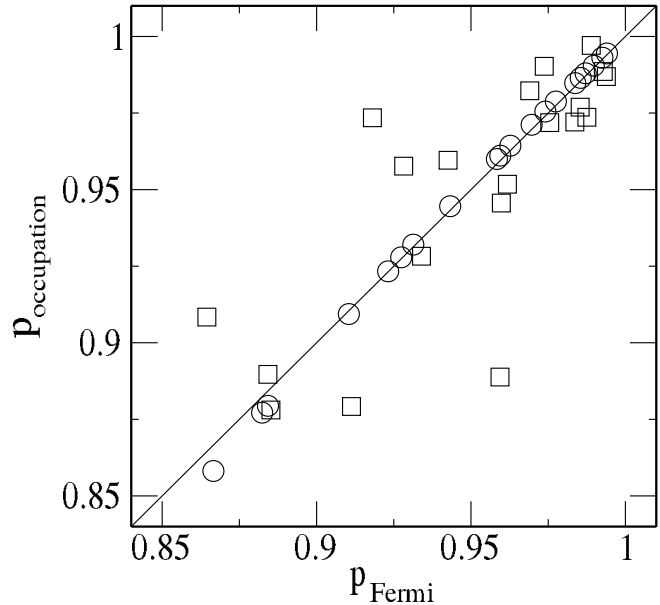}
\caption{\label{fig_simple_fermi} Comparison of the probability
$p^{\mathrm{ion}}(s)$ for the ion population of site $s$ in a
simple vacancy model (see text) with its Fermi estimate. In one
case (circles) the ion-ion interaction does not depend on the
position of the vacancy, in the other case (squares) a weak
dependence has been taken into account (see text for details).}
\end{center}
\end{figure}

For this model one can first calculate the average energy
$E^{\mathrm{tot}}(s)$ related to site $s$. It is given by
\begin{equation}
E^{\mathrm{tot}}(s)  = E^{\mathrm{net}}(s) + \frac{\sum_{t=1;t\ne
s}^N E^{\mathrm{vac},t}(s) p^{\mathrm{vac},t}}{1-
p^{\mathrm{vac},s}}.
\end{equation}
Via the Boltzmann average this expression implies that the typical residence time is longer than
the typical energy fluctuations via jumps of surrounding particles.
This ansatz is compatible with the simulation results because the fluctuations
of the energy are much faster than typical residence times ($\tau_{fluc} \ll \tau_{res}$.
The probability
that site $s$ is occupied by an ion can be calculated as
\begin{equation}
p^{\mathrm{ion}}(s) = 1- p^{\mathrm{vac},s}.
\end{equation}
In Fig. \ref{fig_simple_fermi} $p^{\mathrm{ion}}(s)$ is compared
to the Fermi distribution, based on the average site--energy
$E^{\mathrm{tot}}(s)$ for the case that $E^{\mathrm{net}}(s)$ is
drawn from a box distribution $[0,4k_B T]$. Without the additional
contribution of $E^{\mathrm{vac},t}(s)$ the Fermi distribution is
indeed very well reproduced. Actually, for just a single vacancy
the Fermi distribution is only an approximation but the difference
to the true distribution is small.  The situation dramatically
changes when the additional contributions $E^{\mathrm{vac},t}(s)$
are taken into account. They are considered as random numbers,
drawn from a narrower box distribution $[0,0.5 k_B T]$. One can
clearly see that now significant deviations from the
Fermi distribution are observed. Thus, even a small interaction
among the ions significantly invalidates the applicability of the
Fermi distribution and renders averaged total site--energies
unsuitable for the prediction of the  population of that site.

\section{Discussion}

The site--energies that are the basis for this investigation of
the energy landscape for ionic transport can be determined with
good precision as averages from long molecular dynamics
simulations. While there are fast fluctuations, the mean value of
the site--energy is stable on the timescale of the typical
residence of alkali ions in a site.  The observed stability of the
site--energies and the absence of relevant relaxations on longer
time scales, i.e. on the scale of ionic diffusion or beyond, are
necessary conditions for our analysis.

A description of the energy landscape through a distribution of static
site--energies, as it is used in most hopping models, is well
justified by these results.
Our analysis demonstrates that
fundamental properties of the energy landscape assumed by these models
depend on the Coulomb interaction among the cations.
Strong correlations in the network energies of neighbor sites
are canceled out by the cation interaction to yield a nearly random landscape.
Both the spatial correlations in $E^{\mathrm{net}}$ and its
anticorrelation with $E^{\mathrm{cat}}$ are reproduced by the
counter ion model\cite{knoedler:1996} by using full Coulomb
interactions between mobile cations and fixed anions on a lattice.
Arguably, any
realistic energy landscape for a hopping model must at least implicitly 
incorporate effects of the cation interaction.
One further example is the
introduction of spatial correlations into the energy landscape of a
model designed to reproduce the internal friction of mixed--alkali
glasses.\cite{peibst:2005}

Both observed correlations in the landscape can be tentatively explained
by an argument similar to that used by Greaves in favor of conduction
channels.\cite{greaves:1985}
He stated that any non--bridging oxygen (NBO) in the structure has to be
shared between several cations in order to fulfill all coordination
requirements. Aided by the
long range of the Coulomb interaction, the presence of an NBO can thus
influence the energies of several neighboring sites alike.
Higher numbers of available NBOs will similarly favor not a single site,
but a region of the system. But such a region will be energetically
favorable to the cations only just until their mutual repulsion balances
the effect of the structure, giving rise to the observed anticorrelation.

The correlations between neighboring sites and between the
different contributions at one site are in turn essential for features
of the landscape that directly affect the dynamics.
While barrier heights for jumps between sites were not
determined, the difference between two site--energies gives a
lower bound for the barrier for the transition between them.
Although the correlations in $E^{\mathrm{net}}$ limit the
occurring differences, the additional smoothing of the landscape
by the anticorrelated $E^{\mathrm{cat}}$ is necessary to reach
values as small as 0.33 eV for \glas{L}{2}. Interestingly, this value
is significantly smaller than the macroscopic activation energy of 0.66 eV.
Two reasons may play a role.
One may expect that the barriers between the sites also make a significant
contribution to the macroscopic activation energy.
Moreover we have seen that the local dynamics is only weakly correlated
with the local energy.

In Fig. \ref{fig_timefit} the correlations between site--energies
and residence times are obscured by scatter that dominates the results for
most sites.
The larger values of $E^{\mathrm{net}}$ observed for slow sites
suggest a low density of non-bridging oxygen
(NBOs) atoms and, correspondingly, a high density of bridging
oxygen (BOs) nearby. This is fully compatible with the previous observation
that sites with long residence times are surrounded by a larger number of
BOs\cite{lammert:2004}. The physical reason is that the jump process of an
ion is supplemented by an instantaneous local door-opening effect of oxygen atoms
\cite{sunyer:2003b,kunow:2005}. In case where the neighborhood
contains BOs the door-opening effect is reduced, giving rise to a
longer residence time. This structural trend is however only sufficient
to indicate very slow sites,
just as the network energies support only a general
relationship, but no clear prediction of the dynamics at any individual site.
The even weaker dependence of the residence times on $E^{\mathrm{tot}}$
indicates that the average site energies do not capture all effects of
the added cation interaction on the local dynamics.

But also considering equilibrium properties the full site--energies,
with the cation interaction included, have still proven insufficient
to determine the behavior of ions at the individual sites.
The most fundamental test for the importance of
the site--energies in our systems is shown in Fig.
\ref{fig_fermi}, where the influence of the site--energies on
their occupation probability is investigated. The relative
occupation of the sites deviates fundamentally from the expected
Fermi dependence.
These observations clearly indicate that average single--site--energies,
as defined here (thus, taking into account the long-range
interaction) and used in many of the models of ion conduction, do
not contain the complete information.

The simple model analyzed in Fig. \ref{fig_simple_fermi} suggests
the reason: Coupling of the site--energies to other, distant parts
of the system makes states of identical local occupation
energetically non-equivalent. In consequence the statistics
yielding e.g. the Fermi distribution are disrupted. In the glass
simulated by the MD model, the fluctuations of the site--energy
are fast compared to the duration of a residence. But for the case
of only a few unoccupied sites the relaxation rate also strongly
depends on the availability of free nearby sites. As jumps happen
with delays much shorter than the total residence times once a
free site is available, the effective timescale on which the
site--energy acts on the occupation becomes comparable to the
timescale of fast energy fluctuations. As a consequence only a weak
correlation between residence time and site--energy is to be expected.

\section{Conclusions}

We have shown that it is possible to describe the energy landscape
for ion conduction in a glass in terms of stable average site--energies.
The analysis of site--energies has established a central influence
of the cation interaction on the basic properties of the energy
landscape.
Yet average site--energies including cation interactions are
still insufficient to predict the ion dynamics directly.
The dependence  of the site--energies on the ionic configuration,
and the low number of free sites, which regulates the
possibility for jumps independently from the energies, are demonstrated
as the likely causes.

A direct description of the dynamics starting from
site--energies will therefore not be able to capture the
correlations among the ions.
Instead the falsification of the single--energy picture
suggests that alternative ways should be explored to build a
description of the ion dynamics from information about the energy landscape.

In particular a treatment in terms of vacancies seems
promising\cite{cormack:2002,lammert:2003,dyre:2003} because due to their
small concentration interaction effects can be expected to be negligible. They offer an
equivalent description of the ion dynamics, without loss of
microscopic detail, because every jump of an ion can be replaced
by a corresponding jump of a vacancy. Work along this line will be
published elsewhere.

\section*{Acknowledgments}
We would like to thank P. Maass for very helpful discussions.
This work was supported in part by SFB 458 and by the Center for
Theoretical Biological Physics sponsored by the NSF(Grant PHY-0822283)
with additional support from NSF-MCB-0543906.


\begin{thebibliography}{10}

\bibitem{baranovskii:1999}
S.~D. Baranovskii and H.~Cordes,
\newblock J. Chem. Phys. {\bf 111}, 7546 (1999).

\bibitem{greaves:1985}
G.~N. Greaves,
\newblock J. Non-Cryst. Solids {\bf 71}, 203 (1985).

\bibitem{greaves:1981}
G.~N. Greaves, A.~Fontaine, P.~Lagarde, D.~Raoux, and S.~J. Gurman,
\newblock Nature {\bf 293}, 611 (1981).

\bibitem{kamijo:1996}
N.~Kamijo, K.~Handa, and N.~Umesaki,
\newblock Materials Transactions, JIM {\bf 37}, 927 (1996).

\bibitem{rouse:1978}
G.~B. Rouse, P.~J. Miller, and W.~M. Risen,
\newblock J. Non-Cryst. Solids {\bf 28}, 193 (1978).

\bibitem{swenson:2001a}
J.~Swenson et~al.,
\newblock Phys. Rev. B {\bf 63}, art. no. 132202 (2001).

\bibitem{balasubramanian:1993}
S.~Balasubramanian and K.~J. Rao,
\newblock J. Phys. Chem. {\bf 97}, 8835 (1993).

\bibitem{habasaki:1995}
J.~Habasaki, I.~Okada, and Y.~Hiwatari,
\newblock J. Non-Cryst. Solids {\bf 183}, 12 (1995).

\bibitem{park:1999}
B.~Park and A.~N. Cormack,
\newblock J. Non-Cryst. Solids {\bf 255}, 112 (1999).

\bibitem{angell:1983}
C.~A. Angell,
\newblock Solid State Ion. {\bf 9--10}, 3 (1983).

\bibitem{peibst:2005}
R.~Peibst, S.~Schott, and P.~Maass,
\newblock Phys. Rev. Lett. {\bf 95}, 115901 (2005).

\bibitem{maass:1999}
P.~Maass,
\newblock J. Non-Cryst. Solids {\bf 255}, 35 (1999).

\bibitem{funke:2004}
K.~Funke and R.~D. Banhatti,
\newblock Solid State Ionics {\bf 169}, 1 (2004).

\bibitem{bunde:1991:2}
A.~Bunde, M.~D. Ingram, P.~Maass, and other,
\newblock J. Phys. A: Math. Gen. {\bf 24}, L881 (1991).

\bibitem{hunt:1997}
A.~G. Hunt,
\newblock J. Non-Cryst. Solids {\bf 220}, 1 (1997).

\bibitem{imre:2006}
A.~W. Imre, S.~V. Divinski, S.~Voss, F.~Berkemeier, and H.~Mehrer,
\newblock Journal Of Non-Crystalline Solids {\bf 352}, 783 (2006).

\bibitem{ingram:2006}
M.~D. Ingram, C.~T. Imrie, and I.~Konidakis,
\newblock Journal Of Non-Crystalline Solids {\bf 352}, 3200 (2006).

\bibitem{kirchheim:2001}
R.~Kirchheim and D.~Paulmann,
\newblock J. Non-Cryst. Solids {\bf 286}, 210 (2001).

\bibitem{lammert:2003}
H.~Lammert, M.~Kunow, and A.~Heuer,
\newblock Phys. Rev. Lett. {\bf 90}, 215901 (2003).

\bibitem{habasaki:2004}
J.~Habasaki and Y.~Hiwatari,
\newblock Phys. Rev. B {\bf 69}, 144207 (2004).

\bibitem{vogel:2004}
M.~Vogel,
\newblock Phys. Rev. B {\bf 70}, 094302 (2004).

\bibitem{lammert:2005}
H.~Lammert and A.~Heuer,
\newblock Phys. Rev. B {\bf 72}, 214202 (2005).

\bibitem{adams:2002}
S.~Adams and J.~Swenson,
\newblock Phys. Chem. Chem. Phys {\bf 4}, 3179 (2002).

\bibitem{habasaki:2007}
J.~Habasaki and K.~L. Ngai,
\newblock Physical Chemistry Chemical Physics {\bf 9}, 4673 (2007).

\bibitem{horbach:2002}
J.~Horbach, W.~Kob, and K.~Binder,
\newblock Phys. Rev. Lett. {\bf 88}, 125502 (2002).

\bibitem{jund:2001}
P.~Jund, W.~Kob, and R.~Jullien,
\newblock Phys. Rev. B {\bf 64}, 134303 (2001).

\bibitem{meyer:2002}
A.~Meyer, H.~Schober, and D.~B. Dingwell,
\newblock Europhys. Lett. {\bf 59}, 708 (2002).

\bibitem{frenkel:2002}
D.~Frenkel and B.~Smit,
\newblock {\em Understanding Molecular Simulations},
\newblock Academic Press, 2nd edition, 2002.

\bibitem{figueirido:1995}
F.~Figueirido, G.~D. {Del Buono}, and R.~M. Levy,
\newblock J. Chem. Phys. {\bf 103}, 6133 (1995).

\bibitem{bansal:1986}
N.~P. Bansal and R.~H. Doremus,
\newblock {\em {H}andbook of {G}lass {P}roperties},
\newblock Academic Press, Orlando, 1986.

\bibitem{habasaki:1992}
J.~Habasaki and I.~Okada,
\newblock Molec. Simul. {\bf 9}, 319 (1992).

\bibitem{hoover:1985}
W.~G. Hoover,
\newblock Phys. Rev. A {\bf 31}, 1695 (1985).

\bibitem{balasubramanian:1995}
S.~Balasubramanian and K.~J. Rao,
\newblock J. Non-Cryst. Solids {\bf 181}, 157 (1995).

\bibitem{habasaki:1996}
J.~Habasaki, I.~Okada, and Y.~Hiwatari,
\newblock J. Non-Cryst. Solids {\bf 208}, 181 (1996).

\bibitem{sunyer:2002}
E.~Sunyer, P.~Jund, and R.~Jullien,
\newblock Phys. Rev. B {\bf 65}, 214203 (2002).

\bibitem{gusev:1991}
A.~A. Gusev and U.~W. Suter,
\newblock Phys. Rev. A {\bf 43}, 6488 (1991).

\bibitem{knoedler:1996}
D.~Kn\"odler, P.~Pendzig, and W.~Dieterich,
\newblock Solid State Ionics {\bf 86--88}, 29 (1996).

\bibitem{lammert:2004}
H.~Lammert and A.~Heuer,
\newblock Phys. Rev. B {\bf 70}, 024204 (2004).

\bibitem{kunow:2005}
M.~Kunow and A.~Heuer,
\newblock Phys. Chem. Chem. Phys {\bf 7}, 2131 (2005).

\bibitem{sunyer:2003b}
E.~Sunyer, P.~Jund, and R.~Jullien,
\newblock J. Phys.: Condens. Matter {\bf 15}, L431 (2003).

\bibitem{cormack:2002}
A.~N. Cormack, J.~Du, and T.~R. Zeitler,
\newblock Phys. Chem. Chem. Phys. {\bf 4}, 3193 (2002).

\bibitem{dyre:2003}
J.~C. Dyre,
\newblock J. Non-Cryst. Solids {\bf 324}, 192 (2003).

\end{thebibliography}

\end{document}